\documentstyle[aps,psfig,multicol,twoside]{revtex}

\begin{document}



\title{Measurements of the instantaneous velocity difference and local
velocity with a fiber-optic coupler}

\author{S. H. Yao}
\address{Department of Physics, Oklahoma State University, Stillwater, OK 74078}
\author{V. K. Horv\'ath}
\address{Department of Physics and Astronomy, University of Pittsburgh,
Pittsburgh, PA 15260}
\author{P. Tong and B. J. Ackerson}
\address{Department of Physics, Oklahoma State University,
Stillwater, OK 74078}
\author{W. I. Goldburg}
\address{Department of Physics and Astronomy, University of Pittsburgh,
Pittsburgh, PA 15260}

\date{\today}

\maketitle

\vskip-2.0truecm
\hskip12.8truecm

\vskip2.4truecm

\bigskip

\begin{abstract}
  New optical arrangements with two single-mode input fibers and a
fiber-optic coupler are devised to measure the instantaneous velocity
difference and local velocity.  The fibers and the coupler
are polarization-preserving to guarantee a high signal-to-noise ratio.
When the two input fibers are used to collect the
scattered light with the same momentum transfer vector but
from two spatially separated regions in a flow, the obtained signals
interfere when combined via the fiber-optic coupler.  The resultant
light received by a photomultiplier tube contains a cross-beat frequency
proportional to the velocity difference between the two measuring points.
If the two input fibers are used to collect the scattered light from a
common scattering region but with two different momentum transfer vectors,
the resultant light then contains a self-beat frequency proportional
to the local velocity at the measuring point.
The experiment shows that both the cross-beat and self-beat signals are
large and the standard laser Doppler signal processor can be used to
measure the velocity difference and local velocity
in real time.  The new technique will have various applications in the
general area of fluid dynamics.

\vskip\baselineskip
\hskip-\parindent

\end{abstract}

\newsavebox{\boxl}
\savebox{\boxl}{$\ell$}
\begin{multicols}{2}

\section{Introduction}

  Measurements of the velocity difference or the relative velocity,
$\delta{\bf v}({\bf \ell})= {\bf v}({\bf x}+{\bf \ell})-{\bf v}({\bf x})$,
between two spatial points separated by a distance $\ell$ have important
applications in fluid dynamics.  For example, in the study of turbulent
flows one is interested in the scaling behavior of $\delta{\bf v}(\ell)$
over varying distance $\ell$, when $\ell$ is in the inertial range, in
which the kinetic energy cascades at a constant rate without
dissipation.\cite{Frisch95,Sreeni99}  If the separation $\ell$ is
smaller than the Kolmogorov dissipation length,\cite{Frisch95,Sreeni99}
the measured $\delta{\bf v}(\ell)$ becomes
proportional to the velocity gradient $\partial v/\partial r\simeq
\delta{\bf v}(\ell)/\ell$ (assuming $\ell$ is known), a useful quantity which
is needed to determine the energy dissipation and flow vorticity.  In many
experimental studies of fluid turbulence, one measures the local velocity
as a function of time and then uses Taylor's frozen turbulence assumption
to convert the measured temporal variations into the spatial fluctuations
of the velocity field.\cite{Taylor38}  The frozen
turbulence assumption is valid
only when the mean velocity becomes much larger than the velocity fluctuations.
For isotropic turbulent flows with a small mean velocity, direct measurement
of $\delta{\bf v}(\ell)$ is needed.

  Over the past several years, the present authors and their collaborators
have exploited the technique of homodyne photon correlation spectroscopy
(HPCS) to measure $\delta{\bf v}(\ell)$.\cite{Narayan97,Yixue98}
With the HPCS scheme, small particles seeded in a flowing
fluid are used to scatter the incident laser light.  The scattered light
intensity $I(t)$, which fluctuates because of the motion of the seed
particles, contains Doppler beat frequencies of all particle pairs in the
scattering volume.  For each particle pair separated by a distance $\ell$
(along the beam propagation direction), their beat frequency
is $\Delta \omega_2={\bf q} \cdot \delta {\bf v}(\ell)$, where $\bf q$ is
the momentum transfer
vector.  The magnitude of $\bf q$ is given by $q=(4\pi n/\lambda)\sin
(\theta/2)$, where $\theta$ is the scattering angle, $n$ is the refractive
index of the fluid, and $\lambda$ is the wavelength of the incident light.
Experimentally, the Doppler beat frequency $\Delta \omega_2$ is measured
by the intensity auto-correlation function,\cite{Berne76}
\begin{equation}
g(\tau)=\frac{\langle I(t+\tau)I(t)\rangle}{\langle I(t)\rangle ^{2}}
=1+bG(\tau),
\label{eq1}
\end{equation}
where $b$ ($\le 1$) is an instrumental constant and henceforth we set
$b=1$.  The angle brackets represent a time average over $t$.

  It has been shown that $G(\tau)$ in Eq. (\ref{eq1}) has the
form\cite{Tong88a}
\begin{equation}
G(\tau)=\int_{0}^{L} dr~h(r) \int_{-\infty}^{+\infty}d\delta v~P(\delta 
v,r) \cos (q\delta v \tau),
\label{eq2}
\end{equation}
where $\delta v$ is the component of $\delta {\bf v}$ in the
direction of $\bf{q}$, $P(\delta v,r)$ is the probability density
function (PDF) of $\delta v(r)$, and $h(r) dr$ is the number fraction 
of particle pairs with separation $r$ in the scattering volume.
Equation (\ref{eq2}) states that the light scattered by each pair of
particles contributes a phase factor $\cos(q\tau \delta {\bf v})$ (because
of the Doppler beat) to the correlation function $G(\tau)$, and
$G(\tau)$ is an {\it incoherent} sum of these ensemble averaged
phase factors
over all the particle pairs in the scattering volume.
In many previous experiments,\cite{Tong88a,Pak91,Tong92,Kellay95}
the length $L$ of the scattering volume viewed
by a photodetector was controlled by the width $S$ of a slit in
the collecting optics.

  While it is indeed a powerful tool for the study of turbulent flows,
the HPCS technique has two limitations in its collecting optics and
signal processing.  First, a weighted average over $r$ is required for
$G(\tau)$ because the photodetector receives light from
particle pairs having a range of separations ($0<r<L$).  As a result,
the measured $G(\tau)$ contains information about $\delta{\bf v}(\ell)$
over various length scales up to $L$.  With the single slit arrangement,
the range of $L$ which can be varied in the experiment is limited.
The lower cut-off for $L$ is controlled by the laser beam radius
$\sigma$.  The upper cut-off for $L$ is determined by the coherence
distance (or coherence area) at the detecting surface of the
photo-detector, over which the scattered electric fields are strongly
correlated in space.\cite{Berne76}  When the slit width $S$ becomes too
large, the photodetector sees many temporally fluctuating speckles (or
coherence areas), and consequently fluctuations in the scattered 
intensity $I(t)$ will be averaged out over the range of $q$-values
spanned by the detecting area.

  Recently, we made a new optical arrangement for HPCS,
with which the weighted average over $r$ in Eq. (\ref{eq2}) is no longer
needed and the upper limit for $L$ can be extended
to the coherence length of the laser.  In the experiment,\cite{Yixue98}
two single mode, polarization-maintaining (PM)
fibers are used to collect light with the same polarization and
momentum transfer vector $\bf q$ but
from two spatially separated regions in a flow.  These regions are
illuminated by a single coherent laser beam, so that the collected signals
interfere when combined using a fiber-optic coupler, before being directed
to a photodetector.  With this arrangement, the measured $G(\tau)$
becomes proportional to the Fourier cosine transform of the PDF
$P(\delta v,r)$.

  The second limitation of HPCS is related to signal
processing.  The correlation method is very effective in picking up small
fluctuating signals, but the resulting correlation function $G(\tau)$ is a
time-averaged quantity.  Therefore, the correlation method is not
applicable to unstable flows.  Furthermore, information about the odd
moments of
$P(\delta v,r)$ is lost, because the measured $G(\tau)$ is
a Fourier cosine transform of $P(\delta v,r)$.

  In this paper, we present a further improvement for HPCS, which
is free of the two limitations discussed above.  By combining the new
fiber-optic method with the laser Doppler velocimetry (LDV) electronics,
we are able to measure the instantaneous velocity difference
$\delta{\bf v}(\ell,t)$ and local velocity ${\bf v}({\bf x},t)$ at a high
sampling rate.  With this technique, the statistics of
$\delta{\bf v}(\ell,t)$ and ${\bf v}({\bf x},t)$ are obtained
directly from the time series data.  The new method of measuring
${\bf v}({\bf x},t)$ offers several advantages over the standard LDV.
The remainder of the paper is
organized as follows.  In Section 2 we
describe the experimental methods and setup.  Experimental
results are presented and analyzed in Section 3.  Finally, the work is
summarized in Section 4.

\section{Experimental Methods}
\subsection{Measurement of the velocity difference}

  Figure 1 shows the optical arrangement and the flow cells used in the
experiment.  A similar setup has been described elsewhere,\cite{Yixue98} and
here we mention only some key points.  
\begin{figure}
   \psfull
   \centerline{\psfig{figure=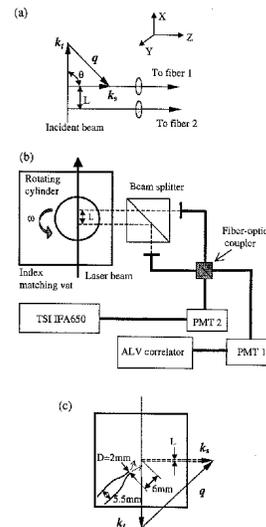,height=7truecm,clip=true}}
   \caption{\narrowtext(a) Scattering geometry for the velocity difference measurement.
$\bf k_i$, incident wave vector; $\bf k_s$, scattered wave vector;
and ${\bf q}={\bf k_s}-{\bf k_i}$.
(b) Experimental setup for the velocity difference
measurement in rigid body rotation.  (c) Flow cell and optical
arrangement for a jet flow. \label{fig1}}
\end{figure}
As shown
in Fig. 1(b), an incident beam from a Nd:YVO$_4$ laser with a power range
of 0.5-2W and a wavelength of $\lambda$=532 nm is directed to a flow cell
by a lens.  With the aid of a large beam splitting cube, two single-mode,
polarization-maintaining (PM) fibers collect the scattered light from
two different spots along the laser beam with a separation $L$.
The two PM fibers are connected to a fiber-optic coupler
(purchased from OZ Optics\cite{oz}), which combines the light from the
two input fibers and split the resultant light evenly into two output fibers.
A graded index lens is placed at each end of the fiber to collimate the
light entering (or exiting from) the fiber core.  Each input fiber
is mounted on a micrometer-controlled translation stage and the
distance $L$ can be adjusted in a range of 0-25 mm in steps of 0.01 mm.
The output fibers of the optical coupler are connected to two photomultiplier
tubes (PMT1 and PMT2).  PMT1 is operated in the digital mode and its
output signal is fed to a digital correlator (ALV-5000).  PMT2 is operated
in the analogue mode and its output signal is fed to a LDV signal
processor (TSI IFA655).  An oscilloscope connected to PMT2 directly views
the analogue signals.  A low-noise preamplifier
(Stanford Research SR560) further amplifies the analogue output of
PMT2 before it goes to the LDV signal processor.

  As shown in Fig. 1(a), the electric fields detected by each input fiber
sum in the coupler and consequently interfere.  In the experiment, we obtain
the beat frequency,
$\Delta \omega_2 ={\bf q} \cdot {\bf v}_1-{\bf q} \cdot {\bf v}_2=
{\bf q} \cdot \delta {\bf v}(L)$, in two different ways.  One way is to
measure the intensity auto-correlation function $g(\tau)$ in
Eq. (\ref{eq1}).  With the ALV correlator, it takes $\sim 1$ minute to collect
the data with an adequate signal-to-noise ratio.  The other way is to
use the LDV signal processor to measure the instantaneous beat frequency
$\Delta \omega_2$, giving velocity differences in real time.
The LDV signal processor is essentially a very fast correlator and
thus requires the beat signal to be large enough so that no signal averaging
is needed.  In the experiment to be discussed below, we use both methods
to analyze the beat signals and compare the results.

  It has been shown that the correlation function $g(\tau)$ has the
form:\cite{Yixue98}
\begin{eqnarray}
g(\tau ) &=& 1+ \frac{I_{1}^2+ I_{2}^2}{(I_{1}+I_{2})^2} G_s(\tau)
+\frac{2 I_{1}I_{2}}{(I_{1}+I_{2})^2} G_c(\tau) \nonumber\\
&=& 1+b_{s}G_{s}(\tau )+b_{c}G_{c}(\tau ),
\label{eq5}
\end{eqnarray}
where $I_1$ and $I_2$ are the light intensities from the two input fibers.
When $I_1=I_2$, one finds $b_s=b_c=0.5$.  If one of the
input fibers is blocked (i.e., $I_2=0$), we have $g(\tau ) = 1+
G_{s}(\tau)$, where $G_{s}(\tau)$ is the self-beat correlation
function for a single fiber.  When the separation L between the two input
fibers is much larger than the spot size viewed by each fiber,
the cross-beat correlation function $G_c(\tau)$ takes the form
\begin{equation}
G_c(\tau) \simeq G_s(\tau) \int_{-\infty}^{+\infty} d\delta v
~P(\delta v) \cos (q\delta v(L)\tau).
\label{eq8}
\end{equation}

  Two flow cells are used in the experiment.  The first one is a
cylindrical cuvette having an inner diameter of 2.45 cm and a height of
5 cm.  The cuvette is top mounted on a geared motor, which produces
smooth rotation with an angular velocity $\omega=2.5 ~rad/s$.  The cell
is filled with 1,2-propylene glycol, whose viscosity is 40 times larger than
that of water.  The whole cell is immersed in a large square
index-matching vat, which is also filled with 1,2-propylene glycol.
The flow field inside the cell is a simple rigid body rotation.  With the
scattering geometry shown in Fig. 1(b), the beat frequency is given by
$\Delta \omega_2=k_s \omega L$ with $k_s=2\pi n/\lambda$.  The sample cell
is seeded with a small amount of polystyrene latex spheres.  For the
correlation measurements, we use small seed particles of $1.0 ~\mu m$
in diameter.   By using the small seed particles, one can have more particles in 
the scattering volume 
even at low seeding densities.  This will reduce the amplitude of the number 
fluctuations caused by a change in the number of particles in each
scattering volume. The particle number fluctuations can
produce incoherent amplitude fluctuations to the scattered light and thus 
introduce an extra 
(additive) decay to $g(\tau)$.\cite{Tong93}  Large particles
$4.75 ~\mu m$ in diameter are used to produce higher scattering intensity
for instantaneous Doppler burst detection.  Because the densities of
the latex particles and
the fluid are closely matched, the particles follow the local flow well
and they do not settle much.

  The second flow cell shown in Fig. 1(c) is used to generate a jet flow
in a $9 cm\times 9 cm$ square vat filled with quiescent water.
The circular nozzle has an outlet 2 mm in diameter and the tube diameter
before the contraction is 5.5 mm.  The nozzle is immersed in the square vat
and a small pump is used to generate a jet flow at a constant flow rate
$0.39~cm^3/s$.  The mean velocity at the nozzle exit is $12.4 ~cm/s$ and
the corresponding Reynolds number is Re=248.  This value of Re is
approximately three times larger than the turbulent transition Reynolds
number, $Re_c\simeq 80$, for a round jet.\cite{Daily66}
Because of the large area
contraction (7.6:1), the bulk part of the velocity
profile at the nozzle exit is flat.  This uniform velocity profile vanishes
quickly, and a Gaussian-like velocity profile is developed in the
downstream region, 2-20 diameters away from the nozzle exit.\cite{White91}
As shown in Fig. 1(c), the direction of the momentum transfer vector $\bf q$
(and hence the measured velocity difference) is parallel to the jet
flow direction, but the separation L is at an angle of $45^o$ to
that direction.

\subsection{Measurement of the local velocity}

  In the measurement of the velocity difference, we use two input fibers to
collect the scattered light with the same $\bf q$ (i.e. at the same
scattering angle) but from two spatially separated regions in the flow.
We now show that with a different optical arrangement, the fiber-optic method
can also be used to measure the instantaneous local velocity
${\bf v}({\bf x},t)$.
Instead of collecting light from two spatially separated regions with the
same $\bf q$, we use the two input fibers to collect light from a common
scattering volume in the flow but with two different momentum transfer
vectors $\bf q_1$ and $\bf q_2$ (i.e. at two different scattering angles).
Figure 2(a) shows the schematic diagram of the scattering geometry.
The collected signals at the two scattering angles
are combined by a fiber-optic coupler, and the resultant light
is modulated at the Doppler beat frequency:\cite{Durst71}
$\Delta \omega_1 ={\bf q_1}\cdot {\bf v} -{\bf q_2}\cdot
{\bf v} =\Delta {\bf q}\cdot {\bf v}({\bf x})$,
where $\Delta {\bf q} =\bf q_1-\bf q_2$.  The magnitude of
$\Delta {\bf q}$ is given by
$\Delta q=(4\pi n/\lambda) \sin(\alpha/2)$, with $\alpha$ being the
acceptance angle between the two fibers.

  The principle of using the scattered light at two different scattering
angles to measure the local velocity has been demonstrated many years
ago.\cite{Durst71}  What is new here is the use of the
fiber-optic coupler for optical mixing.  The fiber-optic technique
simplifies the standard LDV optics considerably.  As shown in Fig. 2(b),
in the standard LDV arrangement the two incident laser beams form
interference fringes
at the focal point.  When a seed particle traverses the focal region,
the light scattered by the particle is modulated by the interference fringes
with a frequency,\cite{Drain80}
$\Delta \omega_1 ={\bf q_1}\cdot {\bf v}-{\bf q_2}\cdot {\bf v}
=\Delta {\bf q}\cdot {\bf v}({\bf x})$.
The magnitude of $\Delta {\bf q}$ has the same expression
$\Delta q=(4\pi n/\lambda) \sin(\alpha/2)$ as shown in the above, but
$\alpha$ now becomes the angle between the two incident laser beams.
The main difference between the standard LDV and the new fiber-optic method
is that the former employs two incident laser beams and a receiving fiber
[Fig. 2(b)], while the latter uses only one incident laser beam and two
optical fibers to measure each velocity component [Fig. 2(a)].
Consequently, the beat frequency $\Delta \omega_1$ in
Fig. 2(a) is independent of the direction ($\bf k_i$) of the incident
laser beam, whereas in Fig. 2(b) it is independent of the
direction ($\bf k_s$) of the receiving fiber.
\begin{figure}
   \psfull
   \centerline{\psfig{figure=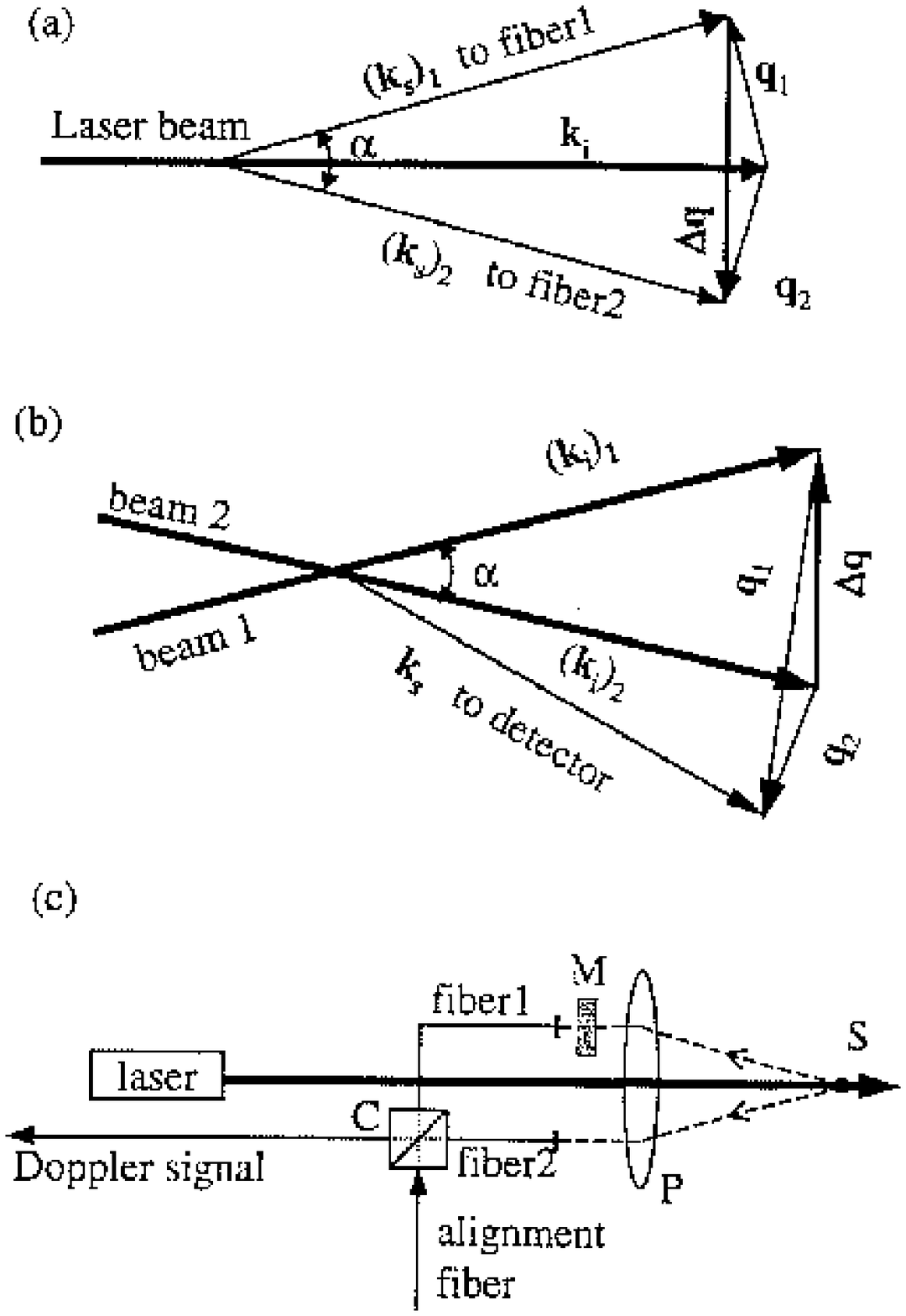,height=6truecm,clip=true}}
\caption{\narrowtext(a) One-beam scattering geometry for the local velocity
measurement. $\bf k_i$, incident wave vector; $({\bf k_s})_1$ and
$({\bf k_s})_2$, two scattered wave vectors;
${\bf q}_1=({\bf k_s})_1-{\bf k_i}$; ${\bf q}_2=({\bf k_s})_2-{\bf k_i}$;
$\Delta {\bf q}={\bf q}_2-{\bf q}_1$.
(b) Two-beam scattering geometry for the local velocity measurement.
$({\bf k_i})_1$ and $({\bf k_i})_2$, two incident wave vectors;
$\bf k_s$, scattered wave vector;
${\bf q}_1={\bf k_s}-({\bf k_i})_1$; ${\bf q}_2={\bf k_s}-({\bf k_i})_2$;
$\Delta {\bf q}={\bf q}_2-{\bf q}_1$.
(c) Schematic diagram of a one-beam probe for the local velocity
measurement. S, measuring point; P, lens; M, frequency modulator; C,
fiber-optic coupler.}
\label{fig2}
\end{figure}

  With the one-beam scheme, one can design various optical probes
for the local velocity measurement.  Figure 2(c) shows an example,
which would replace the commercial LDV probe by reversing the roles of the
transmitter and receiver.  The two input fibers aim at the same
measuring point S through a lens P, which also collects the back-scattered
light from S.
The frequency modulator M shifts the frequency of the light collected by
the input fiber 1, before it is combined via the fiber-optic coupler C with
the light collected by the input fiber 2.  The resultant light from an
output fiber of the coupler contains the beat frequency
$\Delta \omega_1$ and is fed to a photodetector.  Because the measured
$\Delta \omega_1$ is always a positive number, one cannot tell the sign of
the local velocity when zero velocity corresponds to a zero beat frequency.
The frequency shift by the modulator M causes the interference fringes to
move in one direction (normal to the fringes) and thus introduces an extra
shift frequency to the measured beat frequency.  This allows us to measure
very small velocities
and to determine the sign of the measured local velocity relative to the
direction of the fringe motion (which is known).\cite{Drain80}  The
other output fiber of the
coupler can be used as an alignment fiber, when it is connected to a
small He-Ne laser.  With the reversed He-Ne light coming out of
the input fibers, one can directly observe the scattering volume viewed
by each input fiber and align the fibers in such a way that only
the scattered light from the same measuring point S is collected.

  The one-beam probe has several advantages over
the usual two-beam probes.  To measure two orthogonal velocity components
in the plane perpendicular to the
incident laser beam, one only needs to add an extra pair of optical
fibers and a coupler and arrange them
in the plane perpendicular to that shown in Fig. 2(c)
(i.e., rotate the two-fiber plane shown in Fig. 2(c) by $90^o$).
With this arrangement, the four input fibers collect the scattered
light from
the same scattering volume but in two orthogonal scattering planes.
Because only one laser beam is needed for optical mixing, a small
single-frequency diode
laser, rather than a large argon ion laser, is sufficient for the
coherent light source.
In addition, the one-beam arrangement does not need any optics for
color and beam separations and thus can reduce the manufacturing cost
considerably.  With the one-beam scheme, one can make small
invasive or non-invasive probes consisting of only three thin fibers.
One can also make a self-sustained probe containing all
necessary optical fibers, couplers, photodetectors, and a signle-frequency
diode laser.

\section{Results and Discussion}

\subsection{Velocity difference measurements}

  We first discuss the measurements of the velocity difference in rigid
body rotation.  Figure 3(a) shows a sample oscilloscope trace of
the analogue output from PMT2 when the separation L=1.0 mm.  The signal is
amplified 1250 times and band-pass-filtered with a frequency range
of 1-10 kHz.  This oscilloscope trace strongly resembles the burst
signals in the standard LDV.  The only difference is that the signal shown
in Fig. 3(a) results from the beating of two moving particles separated by
a distance L.  Figure 3(a) thus demonstrates that the beat signal between
the two moving particles
is large enough that a standard LDV signal processor can be used to
measure the instantaneous velocity difference $\delta v(\ell,t)$ in real time.

  Figure 3(b) shows the measured beat frequency $\Delta \omega_2$
as a function of separation L.  The circles are obtained from the
oscilloscope trace and the triangles are obtained from the intensity
correlation function $g(\tau)$.  The two measurements agree well
with each other.  The solid line is the linear fit
$\Delta \omega_2=41.78 L ~(10^3 ~rad/s)$, which is in good agreement
with the theoretical calculation
$\Delta \omega_2=k_s \omega L= 42.28 L ~(10^3 ~rad/s)$.  This result also
agrees with the previous measurements by Du et al.\cite{Yixue98}
Because $\Delta \omega_2$ increases with L, one needs to increase the laser
intensity at large values of L in order to resolve $\Delta \omega_2$.
The average photon count rate should be at least twice the measured
beat frequency.
\begin{figure}
   \psfull
   \centerline{\psfig{figure=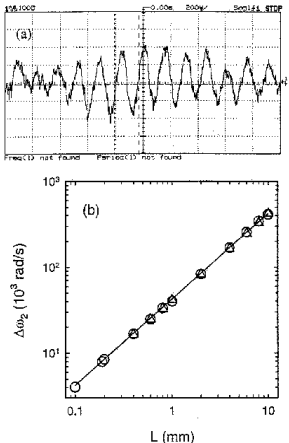,height=6truecm,clip=true}}
\caption{\narrowtext(a) Oscilloscope trace of a typical beat burst between
two moving particles separated by a distance L=1.0 mm. The signal
is obtained in rigid body rotation.  (b) Measured beat frequency
$\Delta \omega_2$ as a function of separation L.  The circles are
obtained from the oscilloscope trace and
the triangles are obtained from the intensity correlation function
$g(\tau)$.  The solid line shows a linear fit to the data points.}
\label{fig3}
\end{figure}

  We now discuss the time series measurements of $\delta v(L,t)$ in a jet
flow using the LDV signal processor.  The jet flow has significant velocity
fluctuations as compared with the laminar rigid body rotation.
The measuring point is in the developing region of the jet flow, 3 diameters
away from the nozzle exit and is slightly off the centerline of
the jet flow.  Figure 4 shows the measured histogram $P(\delta v)$ of
the velocity difference $\delta v(L,t)$ in the jet flow, when
the separation L is fixed at L=0.5 mm (circles) and L=0.8 mm (squares),
respectively.  It is seen that the measured $P(\delta v)$ has a dominant
peak and its position changes with L.  Because $\delta v(L,t)$ increases with
L, the peak position moves to the right for the larger value of L.  The
solid curve in Fig. 4 is a Gaussian fit to the data points with L=0.5 mm.
The obtained mean value of $\delta v(L,t)$ is
$\langle\delta v\rangle =1.87~cm/s$ and the standard deviation
$\sigma= 0.171~cm/s$.  At L=0.8 mm, the measured $P(\delta v)$ peaks
at the value $\langle\delta v\rangle=2.73 ~cm/s$.
\begin{figure}
   \psfull
   \centerline{\psfig{figure=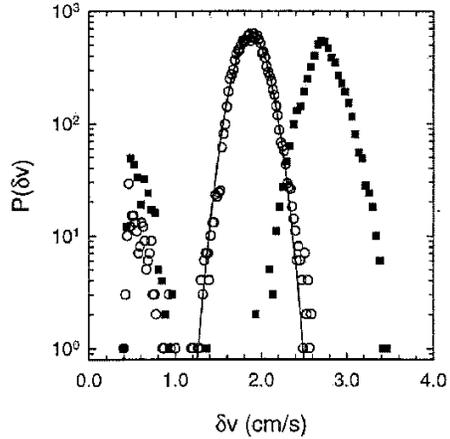,height=6truecm,clip=true}}
\caption{\narrowtext Measured histogram $P(\delta v)$ of the velocity difference
$\delta v(L,t)$ in the jet flow.  The values of L are: L=0.5 mm (circles)
and 0.8 mm (squares).  The solid curve is a Gaussian fit to the circles.}
\label{fig4}
\end{figure}

  In the above discussion, we have assumed that each input fiber sees only
one particle at a given time and the beat signal comes from
two moving particles separated by a distance L.  In fact, when the seeding
density is high, each input fiber may see more than one particle at a given
time.  The scattered light from these particles can also beat and generate a
self-beat frequency proportional to $\delta v(\ell_0)$, where
$\ell_0\simeq 0.15~mm$ is the laser spot size viewed by each input
fiber.\cite{Yixue98}  The self-beating gives rise to a small peak
on the left side of the measured $P(\delta v)$.  Note that the peak position
is independent of L, because $\delta v(\ell_0)$ is determined only by
$\ell_0$, which is the same for both measurements.  It is seen from
Fig. 4 that the cross-beating is dominant over the self-beating under the
current experimental condition.

  The intensity correlation function $g(\tau)$ is also used to analyze the
beat signal.  In the experiment, we measure the histogram $P(\delta v)$
and $g(\tau)$ simultaneously, so that Eq. (\ref{eq8}) can be examined
in details.  Figure 5 shows the measured $g(\tau)-1$ (circles) as a
function of delay time $\tau$ at L=0.5 mm.  The squares are the self-beat
correlation function $G_s(\tau)$ obtained when one of the input fibers
is blocked.  As shown in Fig. 4, the measured $P(\delta v)$ has a Gaussian
form and thus the integration in Eq. (\ref{eq8}) can be carried out.
The final form of $g(\tau)$ becomes
\begin{equation}
g(\tau) = 1+G_s(\tau) \left [b_s+b_c\cos[q\langle\delta v\rangle\tau]
e^{-(q\sigma\tau)^2/2}\right ].
\label{eq9}
\end{equation}
The solid curve in Fig. 5 is a plot of Eq. (\ref{eq9}) with $b_s=0.5$ and
$b_c=0.13$.  The values of $\langle\delta v\rangle$ and $\sigma$ used in
the plot are obtained from the Gaussian fit shown in Fig. 4.
It is seen that the calculation is in good agreement with the measured
$g(\tau)$.

  The fitted value $b_s=0.5$ agrees with the expected value at
$I_1=I_2$.  The value of $b_c$ would be 0.5 if the collected signals
from the two input fibers were fully coherent and the fiber-optic
coupler mixed them perfectly.  The fact that the fitted value of $b_c$
is smaller than 0.5 indicates that the collected signals are not fully
correlated.  This is caused partially by the fact that in the present
experiment the scattered light suffers relatively large number fluctuations
resulting from a changing number of particles in the scattering volume.
These number fluctuations produce incoherent
amplitude fluctuations to the scattered light and thus introduce
an extra (additive) decay to $g(\tau)$.\cite{Tong93}
Because the beam crossing time (proportional to the beam diameter)
is much longer than the Doppler beat time
$1/\Delta \omega_2$ (proportional to the wavelength of the scattered light),
the slow decay due to the number fluctuations can be
readily identified in the measured $g(\tau)$.  This decay
has an amplitude 0.4 and has been subtracted out from the measured
$g(\tau)$ shown in Fig. 5.
\begin{figure}
   \psfull
   \centerline{\psfig{figure=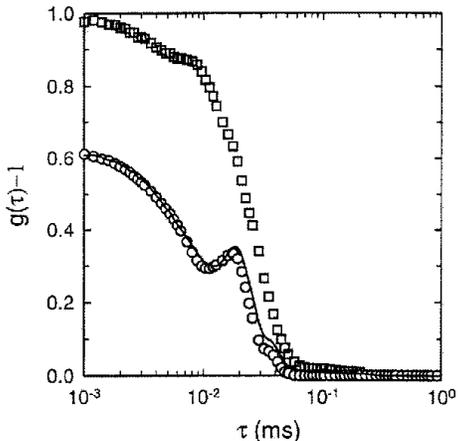,height=6truecm,clip=true}}
\caption{\narrowtext Measured intensity correlation function $g(\tau)-1$ as a function
of delay time $\tau$ at L=0.5 mm (open circles).  The squares are obtained
when one of the input fibers is blocked.  The solid curve is a plot
of Eq. (5).}
\label{fig5}
\end{figure}

  It is shown in Eq. (\ref{eq9}) that to accurately measure the mean velocity
difference $\langle\delta v \rangle$, the beat frequency
$\Delta \omega_2=q\langle\delta v\rangle$ must be larger than the
decay rate $\Gamma_s \simeq q\delta v(\ell_0)$ for $G_s(\tau)$ and also
larger than the decay rate
$\Gamma_c \simeq q\sigma$ resulting from the fluctuations of the velocity
difference.  From the measurements shown in Figs. 4 and 5, we find
$\Gamma_s \simeq 1.33\times 10^5 ~s^{-1}$ and $\Gamma_c \simeq 3.8\times
10^4 ~s^{-1}$, which are indeed smaller than the beat frequency
$\Delta \omega_2\simeq 4.15\times 10^5 ~s^{-1}$.  Because $g(\tau)$
contains a product of $G_s(\tau)$ and $\exp[-(q\sigma\tau)^2/2]$ [see
Eq. (\ref{eq9})], its decay is determined by the faster decaying function.
It is seen from Fig. 5 that the decay of $g(\tau)$ is controlled
by $G_s(\tau)$, which decays faster than $\exp[-(q\sigma\tau)^2/2]$.
It should be noted that in the measurements shown in Fig. 4, the beat
signals are analogue ones and we have used a band-pass filter together with
a LDV signal analyzer to resolve the beat frequency.  Consequently, many
low-frequency self-beat signals are filtered out.  This low-frequency
cut-off is apparent in Fig. 4.  The measurements of $g(\tau)$, on the other
hand, are carried out in the photon counting mode, and
therefore the measured $g(\tau)$ is sensitive to all the self-beat signals
as well as the cross beat signals.  With a simple counting of particle pairs,
we find that the probability for cross beating is only twice larger
than that for the self-beating.

\subsection{Local velocity measurements}

  We now discuss the local velocity measurements using the new optical
arrangement shown in Fig. 2(a).  The velocity measurements are conducted
on a freely suspended flowing soap film driven by gravity.
Details about the apparatus has been described
elsewhere,\cite{Rutgers97,Goldburg98,Horvath00a} and here we
mention only some key points.  $2\%$ solution of detergent and water is
introduced at a constant rate between two long vertical nylon
wires, which are held apart by small hooks.  The width of the channel (i.e.,
the distance between the two nylon wires) is 6.2 cm over a distance of
120 cm.  The measuring point is midway between the vertical wires.
The soap solution is fed, through a valve, onto an apex at
the top of the channel.  The film speed $\bar v$, ranging from 0.5 to
3 m/s, can be adjusted using the valve.  The soap film is approximately
2-6 $\mu m$ in thickness and is seeded with micron-sized latex particles,
which scatter light from a collimated laser beam.  The light source
is an argon-ion laser having a total power of 1W.  The incident laser beam
is oriented perpendicular to the soap film and the scattered light
is collected in the forward direction.

\begin{figure}
   \psfull
   \centerline{\psfig{figure=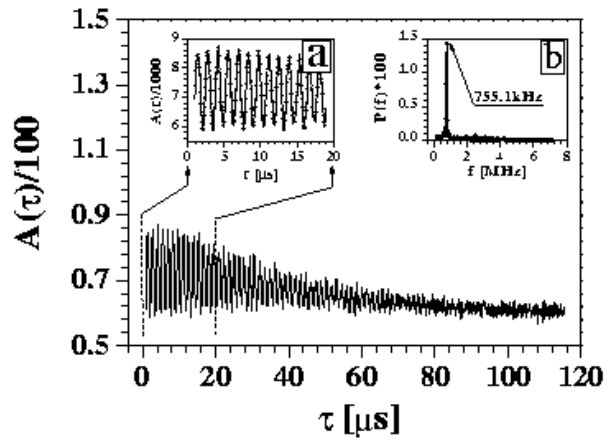,height=6truecm,clip=true}}
\caption{\narrowtext  Measured intensity autocorrelation function
$A(\tau)$ as a function of delay time $\tau$ with the measuring time
T=30 ms.  The inset (a) shows
an enlarged portion of $A(\tau)$ for small values of $\tau$ up to $\tau=20
~\mu s$.  The inset (b) shows the frequency power spectrum $P(f)$ of the
measured $A(\tau)$.}
\label{fig6}
\end{figure}
  To measure the rapidly changing beat frequency $\Delta \omega_1$,
we build a fast digital correlator board for PC.\cite{Horvath00b}
With a fast sampling rate $f_s$, the plug-in correlator board records the
time-varying intensity I(t) (number of TTL pulses from the photomultiplier
tube per sample time) over a short period of time T and then
calculates the (unnormalized) intensity autocorrelation function,
$A(\tau)=\langle I(t+\tau)I(t)\rangle$.
Figure 6 shows an example of the measured $A(\tau)$ as a function of delay
time $\tau$ with T = 30 ms and $f_s=14.32$ MHz.  Because the
burst signal I(t) is a periodic function of t, the measured $A(\tau)$
becomes an
oscillatory function of $\tau$.  The frequency of the oscillation apparent
in the inset (a) is the beat frequency $\Delta \omega_1$.
The amplitude of the oscillation decays at large $\tau$.
The inset (b) shows the power spectrum $P(f)$ of the measured
$A(\tau)$; it reveals a dominant peak at 755.1 kHz.  The power spectrum
is obtained using a fast Fourier transform (FFT) program.\cite{Press92}

\begin{figure}
   \psfull
   \centerline{\psfig{figure=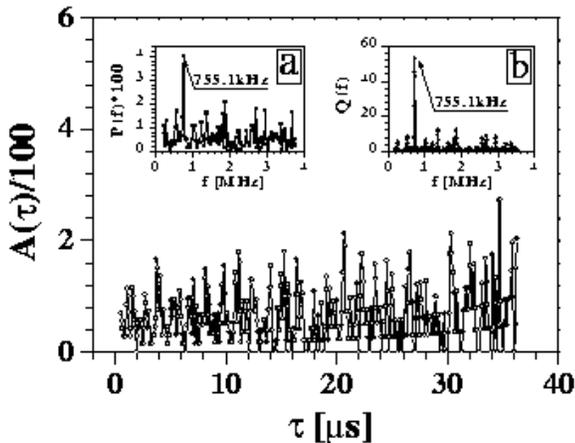,height=6truecm,clip=true}}
\caption{\narrowtext Measured intensity autocorrelation function
$A(\tau)$ as a function of delay time $\tau$ with the measuring time
T=50 $\mu s$.  The inset (a) shows the frequency power spectrum $P(f)$
obtained by FFT.  The inset (b) shows the frequency spectrum $Q(f)$
obtained by the Scargle-Lomb method.}
\label{fig7}
\end{figure}
  To increase the sampling rate of the velocity measurements, one needs to
keep the measuring time T for each $A(\tau)$ as short as possible.
The signal-to-noise ratio for $A(\tau)$ decreases with shorter measuring
time T and with lower mean photon count rate, which was
$\sim 1$ MHz in the present experiment.  It is
found that the shortest useful measuring time $T_c$ is roughly
$50~\mu s$.  For this value of T, $A(\tau)$ is quite noisy and
the corresponding peak in $P(f)$ becomes less pronounced [see Fig. 7 and
inset (a)].  It is worth mentioning that if one only wants to know the
periodicity of a function, rather than its actual power at different
frequencies, the Scargle-Lomb method\cite{SL} is a better alternative to
FFT.  This method, which does not require evenly spaced sampling,
compares the measured data with known periodic signals using
the least-square fitting procedure and determines the relevant frequencies
by the goodness of the fit $Q(f)$.  It can even utilize the uneven sampling
to further increase the Nyquist frequency.  As shown in Fig. 7(b), the
Scargle-Lomb
method can still clearly identify the periodicity of the signal, even
when the power spectrum $P(f)$ [Fig. 7(a)] becomes less reliable.
The total time required for the measurement of the characteristic
frequency is less than 1 ms.  Using the correlator board together with
an average speed PC (300 MHz), we are able to conduct accurate measurements
of the local velocity with a sampling rate up to 1 kHz.

\section{Summary}

  We have developed new optical arrangements with two single-mode
input fibers and a fiber-optic coupler to measure the local
velocity ${\bf v}({\bf x})$ and the velocity difference,
$\delta{\bf v}({\bf \ell})= {\bf v}({\bf x}+{\bf \ell})-{\bf v}({\bf x})$,
between two spatial points separated by a distance $\ell$.  The fibers and
the coupler are polarization preserving to guarantee a high signal-to-noise
ratio.
  To measure the velocity difference $\delta{\bf v}({\bf \ell})$, the two
input fibers are used to collect the
scattered light with the same momentum transfer vector $\bf q$ but
from two spatially separated regions in a flow.  These regions are
illuminated by a single coherent laser beam, so that the collected signals
interfere when combined via the fiber-optic coupler.  The resultant
light received by a photomultiplier tube therefore contains the beat
frequency $\Delta \omega_2={\bf q} \cdot \delta {\bf v}(\ell)$.
We analyzed the beat signals using two different
devices and compared the results.  First,
the intensity auto-correlation function $g(\tau)$ was measured using a
digital correlator.  Secondly, a standard LDV signal processor was used to
determine the instantaneous beat frequency $\Delta \omega_2$.
With this device, $\delta{\bf v}({\bf \ell,t})$ can be obtained in real time.
The technique can be further developed to measure one component
of the local flow vorticity vector ${\vec \omega}({\bf x},t) =
{\bf \nabla} \times {\bf v}({\bf x},t)$.\cite{Yao00}

   To measure the instantaneous local velocity itself, one needs only to
reorient the two fibers so that they point to the same scattering volume.
With this optical arrangement, we have three alternatives to measure
a velocity
component.  They employ (i) an analog photodetector and a standard LDV
signal processor (burst detector), (ii) a commercial photon
correlator, such as that made by ALV,
and finally (iii) a home-made digital correlator.  This latter device
completes a velocity measurement in less than 1 ms and is orders of
magnitude cheaper than the other two alternatives.
The new fiber-optic method has several advantages over the standard LDV
and can be used widely in the general area of fluid dynamics.
Because only one laser beam is needed to obtain two velocity components,
a compact single-frequency diode laser can replace a large multi-frequency
argon-ion laser.  By eliminating the color and
beam separation units in
the standard LDV, the one-beam scheme is less costly to implement.
With more optical fiber pairs and couplers, one can carry
out multi-point and multi-component velocity measurements in various
turbulent flows.

\acknowledgments

  We thank M. Lucas and his team for fabricating the scattering
apparatus and J. R. Cressman for his contributions.
The work done at Oklahoma State University was supported
by the National Aeronautics and Space Administration (NASA) Grant
No. NAG3-1852 and also in part by the National Science Foundation (NSF)
Grant No. DMR-9623612. The work done at University of Pittsburgh
was supported by NSF Grant No. DMR-9622699, NASA Grant No.
96-HEDS-01-098, and NATO Grant No. DGE-9804461. VKH acknowledges
the support from the Hungarian OTKA F17310.


\end{multicols}
\end{document}